\documentstyle[eqsecnum,multicol,epsfig,aps,prl,array]{revtex}
\newcommand{\bleq}{\ifpreprintsty
   \else
   \end{multicols}\widetext \vspace*{-3.5ex}{\tiny
   
\noindent\begin{tabular}[t]{c|}
   \parbox{0.493\hsize}{~} \\ \hline \end{tabular}}
      \fi}
\newcommand{\eleq}{\ifpreprintsty
   \else
   {\tiny\hspace*{\fill}\begin{tabular}[t]{|c}\hline
    \parbox{0.49\hsize}{~} \\
    \end{tabular}}\vspace*{-2.5ex}\begin{multicols}{2}
    \narrowtext
    \fi}
\newcommand{\bcols}{\ifpreprintsty\else\begin{multicols}{2} 
\narrowtext\fi}
\newcommand{\ecols}{\ifpreprintsty\else\end{multicols}\fi}

\draft
\widetext
\begin{document}
\title{Correlation between floppy to rigid transitions and non-Arrhenius 
conductivity in glasses} 
\author{M. Malki$^{\dag}$, M. Micoulaut$^{\ddag}$, F. Chaimbault$^{\dag}$, 
Y. Vaills$^{\dag}$}

\address{
$^{\dag}$ CRMHT-CNRS 1D, avenue de la Recherche Scientifique\\
45071 Orl\'eans Cedex 02, France\\
Universit{\'e} d{´}Orl\'eans, 45067 Orl{é}ans Cedex 02, France
\ \\
$^{\ddag}$ Laboratoire de Physique Th\'eorique des Liquides
Universit\'e Pierre et Marie Curie\\  Boite 121
4, Place Jussieu, 75252 Paris Cedex 05, France\\}

\date{\today}
\maketitle
\begin{abstract}
\par
Non-Arrhenius behaviour and fast increase of the ionic conductivity is observed for a number of potassium silicate glasses $(1-x)SiO_2-xK_2O$ with potassium 
oxide concentration larger than a certain value $x=x_c=0.14$.  Recovering of Arrhenius 
behaviour is provided by the annealing that enhances densification.  Conductivity furthermore 
obeys a percolation law with the same critical concentration $x_c$. These various results are the manifestation of the 
floppy or rigid nature of the network and can be analyzed with constraint theory. They underscore the key role played by 
network rigidity for the understanding of conduction and saturation effects in 
glassy electrolytes.
\par
{Pacs:} 61.43Fs-66.10.Ed-66.30.Hs
\end{abstract}
\bcols 
Amorphous electrolytes are a class of materials where cationic or anionic sites are not confined on a specific lattice but are 
essentially free to move throughout the structure. Among those, oxide and chalcogenide fast ionic conducting (FIC) glasses
 show high electrical conductivities\cite{r1,r2} with potential applications for solid state batteries, sensors and 
displays. Recent studies have indeed shown that the dc conductivity of certain classes of sulphide systems at room 
temperature could be increased\cite{r3,Mitkova} up to $10^{-2}\Omega^{-1}.cm^{-1}$. Active research is therefore undertaken to 
make these superionic glasses even more conductive. A natural question that emerges deals with the upper conduction 
limit that could be reached for these systems as saturation effects\cite{PRL96} and departure from Arrhenius behaviour (AB)
in conductivity with increasing temperature have been found to occur. Alternatively, one
 may wonder what kind of chemical elements should be used to avoid the latter. It is generally accepted
that both structural 
and conduction energetic features are involved in the mechanism of superionic conductors and various studies have 
highlighted the role played either by carrier concentration\cite{r5}, mobility or temperature\cite{r6}. Others have 
emphasized the influence of the dynamics\cite{Ngai} or the composition\cite{struct} of these peculiar 
materials to understand their surprisingly high values of conductivity. A basic question is therefore to understand what 
produces the limitation in conductivity \cite{Martin2003}.

From the measurements realized on potassium silicates $(1-x)SiO_2-xK_2O$, we suggest in this Letter that the electrical transport follows
percolative behaviour with alkali concentration and leads to a fast increase of the dc conductivity for $x>x_c=0.14$. 
The latter critical concentration furthermore separates glasses with AB from those displaying non-AB.
We show that annealing and densification of these electrolytes tends to 
reduce the saturation and that densification with annealing may be responsable for the loss of non-Arrhenius variation. 
Finally, we suggest that there 
is a common mechanical origin for the observation of carrier mobility percolation, saturation effects and densification that can be derived directly from
the Phillips-Thorpe constraint theory \cite{JNCS79}. Glasses with 
$x<x_c$ are found to be intrinsically stressed rigid and display AB while those with $x>x_c$ are floppy and show saturation effects and non-AB.
Thus non-AB electrical behaviour intervenes only in floppy glasses. The trends measured here and the critical composition observed
correlate with 
constraint counting algorithms applied on the local structure of the glass. This links for the first the 
elastic nature (floppy, rigid) of the glass network {\em via} its structure, to ionic conductivity and ease of conduction. 
\par 
The glasses were prepared by melting a mixture of $SiO_2$ and $KHCO_3$ powders in a platinum crucible ($200 ~g$ batch) at $1500-1600^oC$ for 
4 hours and were then quenched on a stainless steel. The annealed glasses were obtained by heating the as-quenched melt (termed in the following 
as {\em virgin}) at temperatures close to $T_g$ for 
4 hours followed be a slow cooling down to room temperature. The complex electrical 
conductivity was measured on Pt-metallized discs (1mm thickness and 12 mm diameter) using a Solartron SI 1260 impedancemetre in the frequency range 
1Hz-1MHz from room temperature up to $T_g + 60^oC$. 
The mass densities of virgin and annealed glasses were measured by buoyancy method with an 
accuracy of $0.001~g/cm^3$.
\par
Figure 1 shows the Arrhenius plots of the ionic conductivity for the different virgin potassium silicates. One can first observe 
that glasses at low potassium concentration (e.g. $x=0.05$) display an almost perfect Arrhenius behaviour $\sigma T=\sigma_0exp[-E_A/RT]$
with respect to the temperature while 
those at high concentration exhibit a clear saturation manifested by a 
significant curvature at high temperatures that signals departure from Arrhenius behaviour. 
\begin{figure}
\begin{center}
\epsfig{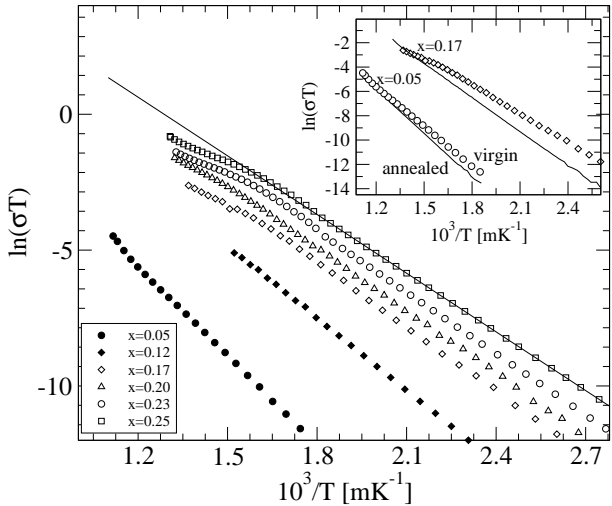}
\end{center}
\caption{Arrhenius plots of virgin potassium silicate glasses with 
changing potassium oxide concentration $x$. The solid line corresponds to a low-temperature Arrhenius fit for the $x=0.25$ 
composition. The insert shows the effect of 
the annealing on the conductivity for two glass compositions. 
Conductivity of virgin glasses: $x=0.05$ (open circles) and 
$x=0.17$ (open diamonds) compared to the one of the annealed glasses 
(solid lines).}
\end{figure}
The present observation has been already reported for a variety of solid electrolytes, 
and it was concluded \cite{PRL96} that ion-ion interactions can be responsible for the latter anomalous behaviour. 
If this would be the case, one would expect to see the saturation emerge in potassium silicates only in the high alkali limit where the proliferation 
of non-bridging anions makes this scenario plausible. Apparently, this seems in contradiction with our results that show already 
non-Arrhenius behaviour for the $x=0.17$ composition. For the latter, $Q^3-Q^3$ bondings that could potentially lead to 
ion-ion interactions are absent \cite{spec} at this concentration. $Q^n$ denotes here a basic $SiO_{4/2}$ tetrahedron having $4-n$ non-bridging oxygens that are only bonded to
a potassium cation. 
\par
The insert of Figure 1 highlights the fact that annealing removes the curvature and brings the conductivity in the glass to Arrhenius behaviour even though 
the absolute value of conductivity decreases with respect to the virgin state. 
This is a feature that has been observed long time ago by Ingram and co-workers \cite{Ingram} for oxysalt chalcogenides involving iodine anions 
and identified with the dynamic temperature dependent restructuring I- sublattice. Densification was claimed to play the key role and it appears that 
this is also the case in the present system as density changes with annealing are relatively small (Figure 3a) for compositions where no departure from Arrhenius 
behaviour is seen. This is also in agreement with the fact that decrease of density 
increases conductivity \cite{PRL96b} in 
fast ion conducting glasses. 
\par
\begin{figure}
\begin{center}
\epsfig{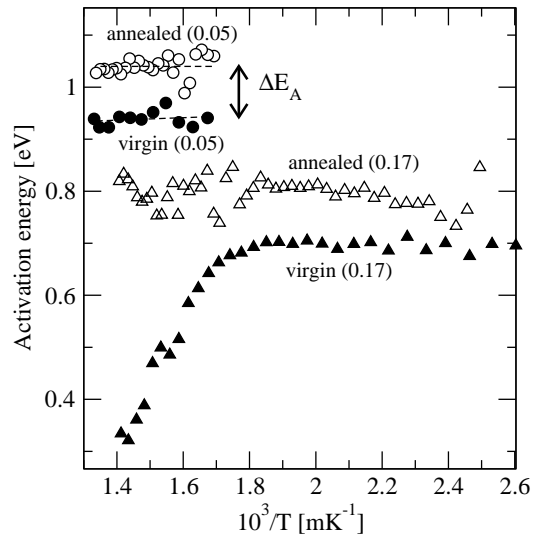}
\end{center}
\caption{Running activation energy with respect to the inverse of temperature for the virgin and annealed 
samples at $x=0.05$ and $x=0.17$. Dotted lines correspond to the Arrhenius fit in Figure 1. $\Delta E_A$
represents the activation energy difference between the annealed and virgin state.}
\end{figure}
Figure 2 shows for two virgin and two annealed potassium glasses ($x=0.05$ and $0.17$) the apparent activation energy that is computed as 
the running slope between adjacent temperature conductivity data points. We note from the figure that in the case of the latter virgin glass, the difference 
in activation energy between the low and the high temperature side is about $0.4~eV$ ($0.7eV$ at $T\simeq 400~K$ i.e. $10^3/T\simeq2.5$, and 
$0.3~eV$ at $T\simeq 700~K$ i.e. $10^3/T\simeq 1.4$)
which is of the order of the low temperature activation energy itself. For comparison, the corresponding reported difference \cite{PRL96} for the 
highest saturated silver superconducting chalcogenide was about $0.15~eV$. The effect of saturation is therefore certainly not weak and does not appear to 
be restricted to chalcogenides \cite{PRL96}. On the other hand, both virgin and annealed ($x=0.05$) glasses display a more or less constant activation 
energy (respectively found as $1.04~eV$
and $0.93~eV$) in harmony with the observed Arrhenius trend in the insert of Fig. 1. From the observations made on both figures, it becomes clear that a dramatic 
change in regime occurs 
in the composition intervall $0.12<x<0.17$. This change is also observable from the low temperature activation 
energy difference between the annealed and virgin state $\Delta E_A=E_A(anneal.)-E_A(virg.)$ which
displays (Fig. 3a) a significant drop in the aforementionned compositional interval.
\par
Lagrange bonding constraint counting (LBCC) that include bond stretching and bond-bending forces, appear to be helpful \cite{JNCS79} in understanding the 
modification of the electrical properties with potassium concentration. Connections of network rigidity with electrical conduction in solid electrolytes 
has been established \cite{Nature} recently. LBCC is also a systematic tool to correlate quantitatively network modification with the glass-forming 
tendency in chalcogenide \cite{PRL2001} and oxide glasses \cite{vaills} and provides the framework to understand the reported results. The present 
system can be seen as composed of a network of N atoms among which $N_r$ are $r$-fold coordinated. The enumeration of mechanical constraints leads to 
$r/2$ bond stretching constraints for an r-fold atom while there are $2r-3$ bond bending constraints \cite{Thorpe83}. The constraints acting on the terminal 
potassium cation and the related non-bridging oxygen are handled as described in Ref. \cite{Boolthorpe} where broken constraints are considered. 
\begin{figure}
\begin{center}
\epsfig{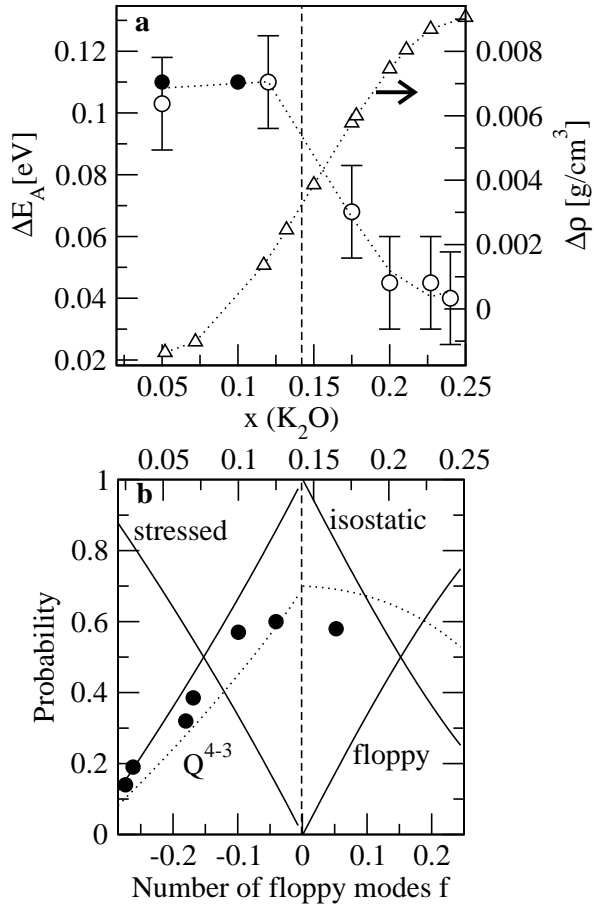}
\end{center}
\caption{a) Low temperature activation energy difference 
$\Delta E_A$ (open circles) between the virgin and annealed samples, together with 
previous results obtained (filled circles, [15]), 
as a function of potassium 
concentration $x$. Right axis shows the corresponding density change $\Delta\rho$
(open triangles) with annealing. b) Relative fraction of $Q^{4-3}$ units 
(filled circles, from [16]) 
together with theoretical prediction of the probability of 
$Q^{4-3}$ units (dotted line) computed 
from size increasing cluster approximations (SICA) [17]. The solid
lines represent the probability of finding stressed rigid, isostatically rigid
and floppy clusters. 
The lower horizontal axis is scaled in atomic number of floppy modes $f$. 
The vertical broken line corresponds to the mean-field rigidity transition 
$x_c=0.14$ where $f=0$.}
\end{figure}
According to this enumeration, a mean-field rigidity transition is expected to occur \cite{Thorpe83} when the number of zero frequency (floppy) modes $f$
vanishes which happens when the number of constraints per atom 
$n_c=5r/2-3$ equals 
the number of degrees of freedom par atom $N_f=3$ in three dimensions. 
In silicates, such a transition occurs at the alkali concentration 
of $x=0.20$ when the cation size 
is small\cite{vaills}. Due to the potassium cation size (cation radius $R_{K^+}=1.33~A$ and $R_{Na^+}=0.95~A$, \cite{Pauling}), 
one expects that a supplementary oxygen angular constraint is broken because the bridging bond angle $Si-O-Si$ in the glass network is found to display wide 
excursions ($135-155^o$) around a mean angle that is much larger \cite{Farnan} than in the sodium analog \cite{anglesod}.
According to this enumeration, a rigid to floppy transition is predicted \cite{calcul} at the concentration of $x_c=0.14(3)$, close to the observed threshold 
observed in Fig. 3a.
The study of the local structure of potassium silicate glasses obtained from NMR investigation provides a supplementary evidence \cite{Sen} 
about the location of the rigid to floppy transition  
because constraint counting can be applied onto the observed $Q^n$ speciation to yield the 
probability of finding stressed rigid and floppy structures in the network (Figure 3b). 
Furthermore, we notice that a cluster composed of a $Q^4$ and $Q^3$ tetrahedron 
(a ``$K_2Si_4O_9$'' like cluster) is optimally constrained (isostatically rigid, satisfying $n_c=3$) \cite{PRB2003} and the corresponding probability of finding the latter is found to be 
maximum \cite{Sen} around the same critical concentration of $x_c=0.14$. Thus at this composition the network is mainly stress free.
For larger concentrations, SICA predicts the growing emergence of 
floppy clusters such as $Q^3-Q^3$ 
bondings.
\par
In a stressed rigid glass, the mobility of the potassium atoms should 
be rather weak because the cations have to
overcome a strong mechanical deformation energy to create a "doorway" 
that allows to move from one
anionic site to another. With the softening of the glass structure produced 
by the addition of potassium oxide, one expects that this energy decreases with
$x$. For instance, in the strong
electrolyte Anderson-Stuart model\cite{Anderson}, the activation energy for
conductivity
$E_A=E_c+E_m$ depends on a Coulombic term $E_c$ controlling the
free carrier rate and a strain term
$E_m$ that contribute to the carrier mobility. The latter can
be thought as the energy required to enlarge the radius of an anionic
site perpendicular to the "doorway" direction of cation displacement. It is thus proportional \cite{Elliot}
to the elastic constant $c_{44}$.
\par
\begin{figure}
\begin{center}
\epsfig{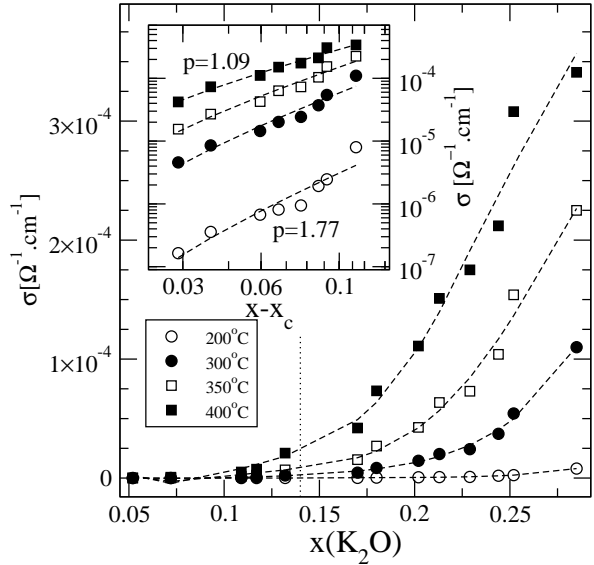}
\end{center}
\caption{\label{logsig}Conductivity of annealed $(1-x)SiO_2-xK_2O$ glasses for 
various temperatures. The insert shows the corresponding logarithmic plot
 against $\log(x-x_c)$. Dotted lines in the insert are the power-law fits.}
\end{figure}

In an ideal floppy glass for which $x>x_c$, this strain energy is 
zero as the related elastic constants ($c_{11}$, $c_{44}$) are found to 
vanish when $f=0$ (or $x=x_c$) \cite{Tersoff,Mike}. This should lead to an 
enhancement of mobility and percolation of floppiness should equal to
percolation of potassium mobility, producing the substantial increase 
of conductivity seen on Figure 4. 
Onset of conductivity in the annealed glasses furthermore follows a 
site percolation law\cite{Scott} of the form: $\sigma(x)/\sigma_0= (x-x_c)^\mu$ that is 
valid for  random or periodic arrays (Insert of Figure \ref{logsig}), with 
powers ranging between $p=1.77$ at $200^oC$ and $1.09$ at $400^oC$. Recently a 
clear experimental correspondence between a previously reported conductivity threshold 
composition \cite{Johari} and the rigidity transition 
composition has been established \cite{boolper} for silver phosphate glasses.
In summary, we have shown that saturation effects and non-Arrhenius behaviour
were taking place in 
the high temperature limit of only certain potassium silicates at compositions 
where the network is floppy. We have pointed out that there was 
also a clear correlation between the onset of conductivity or mobility, effect 
of annealing and the elastic nature of the host network, 
stressed rigid or floppy. We believe that these new 
connections will bring insights and understanding for the improvement 
of amorphous fast ion conductors.
\par
We gratefully acknowledge ongoing discussions with P. Boolchand and  P. Simon. This work was supported by a joint CNRS-NSF grant Int. 
13049. LPTMC is Unit\'e Mixte du CNRS n. 7600, CRMHT Unit\'e Propre du CNRS n. 812.

\ecols

\begin{thebibliography}{100}
\bibitem{calcul} In $(1-x)SiO_2-xK_2O$ glasses, taking the coordination number of Si, O and K respectively of 4, 2 and 1 and assuming that all bonding oxygens 
have broken angular constraints, we obtain: $n_c=(11-14x)/3$ and the number of 
floppy modes vanishes when $x=x_c=0.14(3)$.
\end{thebibliography}
\end{document}